\documentclass[twocolumn,
  nobibnotes,
  aps,
  pra,
  showpacs,
  amsmath,
  amssymb,
  superscriptaddress,
  floatfix,
  10pt]{revtex4-1}

\usepackage{graphicx}
\usepackage{bm}
\usepackage{amsmath}
\usepackage{bbm}

\usepackage{color}

\begin{document}

\title{From weak to strong disorder in Weyl semimetals: Self-consistent Born approximation}%

\author{J. Klier}

\affiliation{Institut f\"ur Nanotechnologie, Karlsruhe Institute of Technology, 76021 Karlsruhe, Germany}

\affiliation{\mbox{Institut f\"ur Theorie der Kondensierten Materie, Karlsruhe Institute of Technology, 76128 Karlsruhe, Germany}}

\author{I.V. Gornyi}

\affiliation{Institut f\"ur Nanotechnologie, Karlsruhe Institute of Technology, 76021 Karlsruhe, Germany}

\affiliation{\mbox{Institut f\"ur Theorie der Kondensierten Materie, Karlsruhe Institute of Technology, 76128 Karlsruhe, Germany}}

\affiliation{A. F. Ioffe Physico-Technical Institute, 194021 St.~Petersburg, Russia}

\author{A.D. Mirlin}

\affiliation{Institut f\"ur Nanotechnologie, Karlsruhe Institute of Technology, 76021 Karlsruhe, Germany}

\affiliation{\mbox{Institut f\"ur Theorie der Kondensierten Materie, Karlsruhe Institute of Technology, 76128 Karlsruhe, Germany}}

\affiliation{Petersburg Nuclear Physics Institute, 188350 St.~Petersburg, Russia}

\affiliation{L.\,D.~Landau Institute for Theoretical Physics RAS, 119334 Moscow, Russia}

\date{\today}
\begin{abstract}
We analyze theoretically the conductivity of Weyl semimetals within the self-consistent Born approximation (SCBA) in the full range of disorder strength, from weak to strong disorder.
In the range of intermediate disorder, we find a critical regime which separates the semimetal and diffusion regimes. While the numerical values of the critical exponents are not expected to be exact within the SCBA, the approach allows us to calculate functional dependences of various observables (density of states, quasiparticle broadening, conductivity) in a closed form. This sheds more light on the qualitative behavior of the conductivity and its universal features in disordered Weyl semimetals.    
In particular, we have found that the vertex corrections in the Kubo formula are of crucial importance in the regime of strong disorder and lead to saturation of the dc conductivity with increasing disorder strength. 
We have also analyzed the evolution of the optical conductivity with increasing disorder strength, including its scaling properties in the critical regime. 

\end{abstract}
\maketitle

\section{Introduction}

In recent years, a major focus in condensed matter physics has been put on three-dimensional Weyl and Dirac semimetals. This interest is motivated by topological phenomena characteristic for these materials and by a deep connection to high-energy (relativistic) quantum field theories. This connection is due to a peculiar band structure with linearly touching bands at certain points in the Brillouin zone, as realized in TaAs~\cite{2015arXiv150204684L,2015arXiv150203807X}, NbAs~\cite{2015arXiv150401350X}, TaP~\cite{Xue1501092}, and NbP~\cite{RIS_5}, which gives rise to such phenomena as chiral anomaly~\cite{reviewVish,0953-8984-27-11-113201,PhysRevLett.111.027201} and emergence of protected Fermi arcs~\cite{PhysRevB.83.205101}. Various experimental observations, such as a giant transversal magnetoresistance~\cite{RIS_1, PhysRevB.92.081306, RIS_5, 2016arXiv161001413C,PhysRevB.91.041203, 2017arXiv170406944R, RIS_6} and a negative longitudinal magnetoresistance~\cite{PhysRevB.88.104412, PhysRevB.89.085126, PhysRevLett.113.247203, Lucas23082016,PhysRevB.91.245157, 2015arXiv150302069G, 0953-8984-27-15-152201, 2015arXiv150606577S, RIS_5,2017arXiv170401038B}, peculiar thermoelectrical effects~\cite{Gooth} and induced superconductivity~\cite{Bachmanne1602983}, promise a huge potential for future applications.   

Transport properties of Weyl semimetals are especially peculiar close to the charge neutrality point. One central aspect of this peculiarity is the appearance of a disordered critical point within the perturbative analysis. This was first pointed out within a mean-field approach in Refs.~\cite{PhysRevB.33.3257} and \cite{PhysRevB.33.3263}; later, the emergence of this critical point was established by a renormalization group (RG) analysis \cite{PhysRevLett.107.196803, PhysRevLett.114.166601,PhysRevB.93.155113} with dimensional regularization and by numerical studies \cite{PhysRevLett.113.026602,PhysRevB.92.115145,PhysRevLett.115.076601,PhysRevB.93.201302,Pix,PhysRevX.8.031076}. Similar results are obtained within the $U(N)$ Gross-Neveu model \cite{PhysRevB.94.220201,PhysRevLett.121.166402}. The self-consistent Born approximation applied for weak and strong disorder in Refs.~\cite{PhysRevB.89.054202,PhysRevB.96.165140} also shows the appearance of the disorder critical point.  Recently, the critical point was also found within the Schwinger-Dyson-Ward approach of 
Ref.~\cite{2018arXiv180809860S}. Related effects of disorder have also been addressed in topological insulators in three and four dimensions, including the limit of the 3D Weyl and Dirac semimetallic phase \cite{PhysRevB.79.045321,PhysRevB.85.155138,PhysRevLett.110.236803,PhysRevLett.112.016402,PhysRevLett.116.066401,PhysRevLett.107.196803}. Beyond the commonly used models of Weyl semimetal with point-like or finite-range disorder, effects of long-range ($1/r^2$) disorder potential have been studied in \cite{PhysRevB.95.075131,PhysRevB.95.014204}. Manifestations of the bulk disorder effects on the surface have been discussed in Ref.~\onlinecite{PhysRevB.96.201401}.

For sufficiently weak disorder (i.e., below the critical strength), the density of states evaluated within the perturbation theory vanishes quadratically as a function of energy around the Weyl point. However, non-perturbative effects were argued to create an exponentially small density of states at the Weyl point. These tails have been considered in Refs. \cite{PhysRevX.6.021042, PhysRevB.94.121107,PhysRevB.95.235101} and \cite{PhysRevB.96.064203}. Analytical calculations of the tails in the density of states were performed in Refs.~\cite{PhysRevB.89.245110} and \cite{PhysRevB.96.174205} for resonant scattering and within a $T$-matrix approach, respectively. Instantons in the replica approach, which are known to produce Lifshitz tails~\cite{Lifshitztails}, have been  calculated in high dimensions in Ref.~\cite{PhysRevB.96.014205}. At the same time, recent works~\cite{2018arXiv180500018B,PhysRevB.98.205134} found that the rare-region effects in Weyl semimetals are very special, and individual local disorder configurations are insufficient to induce a finite density of states. In the strong disorder regime, the density of states is finite at the Weyl point already without invoking exponentially small contributions. 

An interesting question about the behavior of the density of states in the critical regime separating weak- and strong-disorder regimes was addressed in several works. The mean-field approach (controlled by the large number of ``flavors'', $N\gg 1$) results in a square-root low-energy behavior of the density of states~\cite{PhysRevB.33.3257,PhysRevB.33.3263} (see also Sec.~\ref{sec:SCBA} below). Within the RG approach, the density of states also exhibits a power-law dependence on energy at the critical disorder strength. Setting $\epsilon=-1$ in the one-loop RG equations derived 
for $2-\epsilon$ dimensions (i.e., controlled for $|\epsilon|\ll 1$) yields a linear vanishing of the 
density of states, see Refs.~\cite{PhysRevLett.107.196803,PhysRevLett.114.166601,PhysRevB.91.035133,PhysRevB.93.155113}. The second-loop ($\epsilon^2$) contributions to the beta-function were explicitly calculated in Refs. \cite{PhysRevB.93.155113} and \cite{2loop}, implying that the linear behavior is not exact. Both the mean-field and RG approaches are, however, uncontrolled in the physical case of a three-dimensional Weyl semimetal with a number $N$ of Weyl nodes of order unity.  
Most of numerical studies \cite{PhysRevLett.110.236803,PhysRevLett.113.026602,PhysRevB.92.115145,PhysRevLett.115.076601,PhysRevLett.116.066401,PhysRevB.93.201302,Pix} suggest the power-law behavior of the critical density of states which is relatively close to the one-loop RG result. However, the spreading of numerical values (compare, e.g., the correlation-length exponent $\nu = 1.47 \pm 0.03$ in Ref. \cite{PhysRevLett.113.026602} with $\nu = 0.86$ in Ref. \cite{PhysRevLett.110.236803}) reflects a difficulty with extracting the exponents characterizing the true asymptotic behavior.   

In this paper, we use the self-consistent Born approximation (SCBA) which is a microscopic version of the mean-field approach. In general, like other approaches, the SCBA is also not a controlled approximation close to the critical disorder strength. However, the advantage of SCBA compared to other methods is that one can capture analytically the qualitative behavior of various observables and their universal features.  
While the numerical values of the critical exponents are not expected to be exact within the SCBA, the approach allows us to calculate functional dependences of the conductivity of Weyl semimetals on various parameters in a closed form for an arbitrary disorder strength within the unified framework. 

More specifically, we investigate the conductivity in the full range of disorder, from weak to strong (including the critical regime), with the focus on properly including current vertex corrections. Former works included vertex corrections into the consideration for the weak-disorder regime~\cite{PhysRevB.89.014205}. In that regime, the vertex corrections (important for Weyl semimetals even for the pointlike impurity potential) lead to the change in the numerical prefactor in the conductivity. In the present work, we find that the vertex corrections are of particular importance for the strong-disorder regime, where they lead to a saturation of conductivity with increasing disorder strength. To determine this behavior, it is required to consider the full self-consistent equation for the calculation of the density of states and of the real part of self-energy, going beyond the calculation of Refs.~\cite{PhysRevB.33.3257, PhysRevB.33.3263}.

The paper is organized as follows.
In Sec.~\ref{sec:SCBA}, we introduce the model with point-like impurity scattering and discuss the results for the energy dependence self-energy and density of states in the whole range of disorder strength. In Sec.~\ref{sec:SCBAcon}, we calculate the conductivity within the SCBA and analyze its dependence on disorder, temperature, and frequency.  
Our findings are summarized and discussed  in Sec.~\ref{sec:summary}.
Throughout the paper we set $\hbar=c=k_B=1$.

\section{Pointlike impurities in SCBA}
\label{sec:SCBA}

We consider the effects of disorder within SCBA to identify the different phases of disordered Weyl semimetals. 
We analyze the self-energy $\hat{\Sigma}(\textbf{p}, \varepsilon)$ in the (impurity-averaged) Green function generated by impurity scattering,
\begin{equation}\label{green}
\hat{G}(\textbf{p}, \varepsilon)
=\frac{1}{\varepsilon-v\boldsymbol{\sigma}\cdot\textbf{p}-{\hat{\Sigma}}(\textbf{p}, \varepsilon)},
\end{equation}
where the Pauli matrices $\sigma$ operate in pseudospin space and $v$ is the quasiparticle velocity.
The calculations are performed under the assumption that disorder is diagonal in both spin and pseudospin indices, and by neglecting the scattering between different Weyl nodes.
The absence of internode scattering leads to a trivial structure in the node space.
Therefore, the calculated density of states and the conductivities are those per Weyl node.

The pointlike impurity potential has the following form:
\begin{equation}
\hat{V}_{\text{dis}}(\textbf{r})=u_{0}\sum_{i}\delta(\textbf{r}-\textbf{r}_{i})
\mathbbm{1},
\end{equation}
with the unit matrix $\mathbbm{1}$ in the pseudospin space.
For such impurity potential,
the disorder correlator (which is, in general, a rank-four tensor) 
is diagonal and independent of the transferred momentum:
\begin{equation}\label{correlator}
W_{\alpha\gamma\beta\delta}(\textbf{q})=\gamma
\delta_{\alpha\gamma}\delta_{\beta\delta},
\end{equation}
where $\gamma= n_\text{imp}u_0^2$
and $n_\text{imp}$ is the concentration of impurities.

Within the SCBA, the self-energy is given by
\begin{equation}\label{scba}
\Sigma_{\alpha\beta}(\textbf{r},\textbf{r}')=
\int\frac{d^{3}q}{(2\pi)^{3}}W_{\alpha\gamma\beta\delta}(\textbf{q})
e^{i\textbf{q}\cdot(\textbf{r}-\textbf{r}')}G_{\gamma\delta}(\textbf{r},\textbf{r}')
\end{equation}
and is proportional to the unit matrix in the energy-band space for the correlator (\ref{correlator}). 
For the pointlike impurities, Eq.~(\ref{correlator}), the self-energy is momentum-independent, and the self-consistency equation (\ref{scba}) takes the form
\begin{align}\label{selfedef}
\Sigma^{R}(\varepsilon)=
\gamma\int\frac{d^3p}{(2\pi)^3}\left[\frac{1}{\varepsilon-v|p|-\Sigma^R}
+\frac{1}{\varepsilon+v|p|-\Sigma^R}\right].
\end{align}
(The superscript ``R'' indicates the we consider the retarded self-energy.)
Since the integral is divergent at large momenta, we introduce the ultraviolet energy cutoff $\Lambda$
(imposing a hard momentum cutoff at $\Lambda/v$). The integration over the momentum then leads to
\begin{align}\label{self-energy}
\Sigma^R(\varepsilon)=\beta(\varepsilon-\Sigma^R)\left[-1+\frac{\varepsilon-\Sigma^R}{2\Lambda}\ln\left(\frac{\varepsilon-\Sigma^R+\Lambda}{\varepsilon-\Sigma^R-\Lambda}\right)\right],
\end{align}
where we introduced the dimensionless disorder strength  
\begin{align}
\beta=\frac{\gamma\Lambda}{2\pi^2v^3}.
\end{align}

For pointlike impurities, we formally consider arbitrary disorder strengths, including the case of strong disorder, $\beta\gg 1$. As discussed below, for microscopic lattice models, $\beta\gg 1$ can be realized for sufficiently smooth disorder. At the end of Sec.~\ref{sec:SCBAcon}, we describe the generalization of the results obtained for pointlike impurities to the case of smooth disorder. 
The density of states $\rho(\varepsilon)$ is related to the imaginary part of the self-energy as follows:
\begin{equation}
\rho(\varepsilon)=-\frac{1}{\pi \gamma} \text{Im} \Sigma^{R}(\varepsilon).
\label{DOS-def}
\end{equation}
Therefore, within the SCBA, the energy scaling of the density of states is that of the imaginary part of the self-energy.

\begin{figure}
\begin{center}
\includegraphics[width=7.5cm]{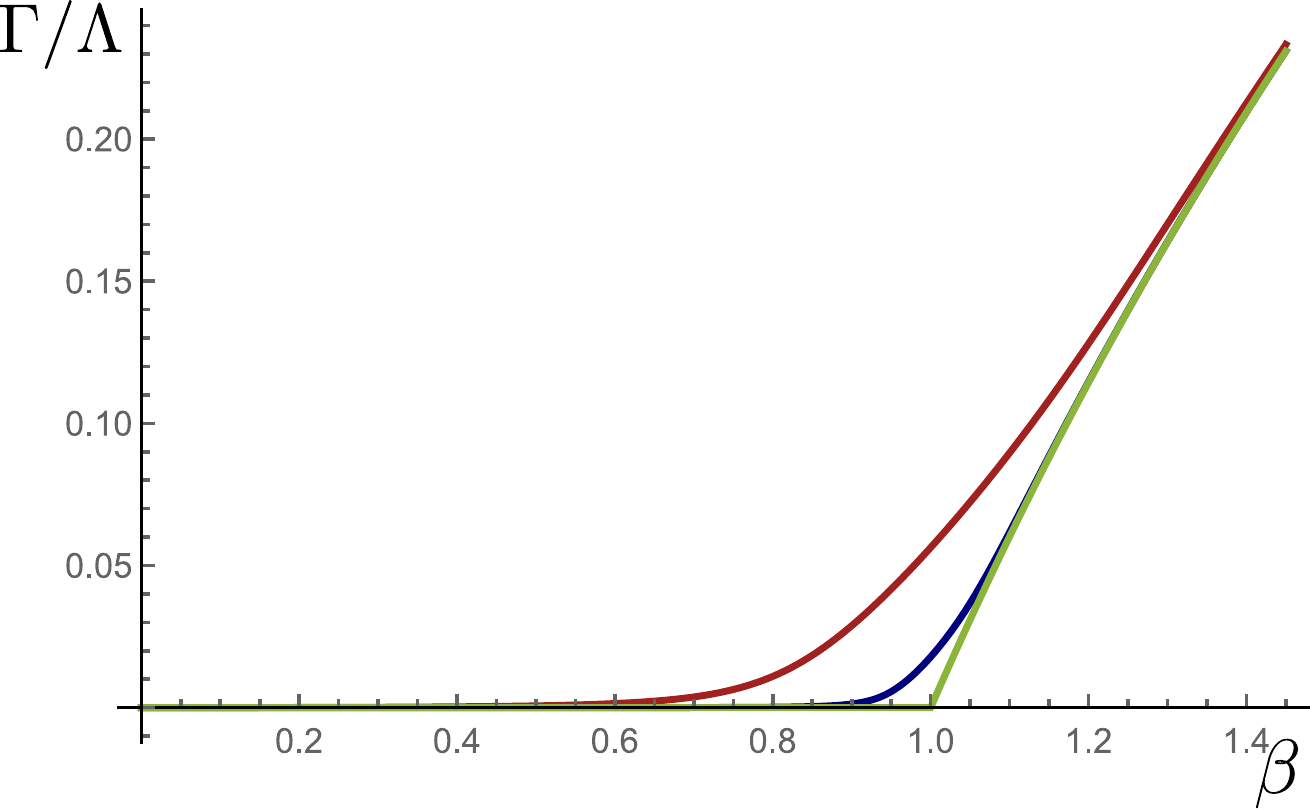} 
\caption{Broadening $\Gamma=-\text{Im}\Sigma^R$ as a function of dimensionless disorder strength $\beta$. Equation~\eqref{imag} is numerically solved for $\varepsilon/\Lambda=0, 10^{-3}, 10^{-2}$ (green, dark blue, and red curves, respectively). The results illustrate analytical asymptotics given by Eqs.~(\ref{strongG}), (\ref{weak}), (\ref{critGamma}).}
\label{fig:imagpart}
\end{center}
\end{figure}

A detailed analysis of Eq.~\eqref{self-energy} is performed in Appendix~\ref{app:self-energy}; here we present and discuss the most salient results. 
We first consider the case of zero energy, $\varepsilon=0$, under the assumption 
$\text{Re}\Sigma^R(\varepsilon=0)=0$ which will be justified later. 
Equation~\eqref{self-energy} gives two solutions for the disorder-induced broadening, 
\begin{equation}
\Gamma=-\text{Im}\Sigma^R.
\end{equation}
The first solution is $\Gamma=0$ and the second is given by the following equation:
\begin{align}\label{gamma0}
\frac{\beta-1}{\beta}=\frac{\Gamma}{\Lambda}\arctan\left(\frac{\Lambda}{\Gamma}\right).
\end{align}
The left-hand side of Eq.~(\ref{gamma0}) exhibits a sign change at $\beta=1$. For $\beta<1$, Eq.~\eqref{gamma0} 
has no physical solution (non-negative $\Gamma$), while for $\beta>1$ a nonzero broadening arises.
This manifests the emergence of the critical point at $\beta=1$. Above the critical disorder strength, a finite density of states is generated. The emergence of this critical point is illustrated by a numerical evaluation of Eq.~\eqref{gamma0} in the full range of disorder in Fig.~\ref{fig:imagpart}. In this figure, the zero-energy broadening is shown by the green curve. For $\beta < 1$ it is zero  as discussed above. For $\beta > 1$ we find
\begin{align}\label{strongG}
\Gamma (\varepsilon =0) =\left\lbrace\begin{array}{cc}
\dfrac{2\Lambda}{\pi}\left(\beta - 1 \right), &\quad \beta-1\ll 1;
\\[0.3cm]
\sqrt{\dfrac{\beta}{3}}\Lambda, &\quad \beta\gg 1.
\end{array}\right.
\end{align}

In the following, we determine the self-energy in the different regimes of disorder for finite energies, 
$\varepsilon>0$. As long as $|\varepsilon-\Sigma^R|\ll\Lambda$, the logarithmic term in 
Eq.~(\ref{self-energy}) can be replaced by a constant $-i\pi$. This results in a quadratic equation for 
the complex quantity $(\varepsilon-\Sigma^R)/\Lambda$ (see Appendix~\ref{app:self-energy}), whose solution reads:
\begin{align}
\text{Re}\Sigma^R&=\varepsilon -\frac{\Lambda|\beta-1|}{\sqrt{2}\pi \beta} 
\sqrt{\sqrt{1+\left[\frac{2\pi \varepsilon}{(\beta-1)^2\Lambda}\right]^2}-1},
\label{ReSigma-sqrt}\\
\Gamma&=\frac{\Lambda(\beta-1)}{\pi\beta}\notag\\
&\quad\ + 
\frac{\Lambda|\beta-1|}{\sqrt{2}\pi\beta}
\sqrt{\sqrt{1+\left[\frac{2\pi \varepsilon}{(\beta-1)^2\Lambda}\right]^2}+1}.
\label{ImSigma-sqrt}
\end{align}
These expressions should be contrasted with the results obtained in Ref.~\cite{PhysRevB.89.054202},
where the inner square roots were in effect expanded in $\varepsilon$. We will see that this approximation is not valid at criticality. Indeed, the behavior of the self-energy is governed by the parameter $(\beta-1)^2\Lambda/|\varepsilon|$. When this parameter is large (i.e., away from the critical point $\beta=1$), one can expand Eqs.~\eqref{ReSigma-sqrt} and \eqref{ImSigma-sqrt} with respect to $\varepsilon/\Lambda$. For $\beta < 1$ this yields 
\begin{align}
\label{weak}
\text{Re}\Sigma^R\simeq-\frac{\beta}{1-\beta}\varepsilon,\qquad
\Gamma\simeq\frac{\pi\beta\varepsilon^2}{2(1-\beta)^3\Lambda}.
\end{align}
The density of states in this regime of weak disorder (or low energies) reads: 
\begin{equation}
\rho(\varepsilon)\simeq \frac{\varepsilon^2}{2\pi^2 v^3(1-\beta)^3}, \qquad |\epsilon|\ll (\beta-1)^2\Lambda.
\label{DOS-weak}
\end{equation}

For critical disorder, $\beta=1$, the self-energy can be written as 
\begin{align}\label{crit}
\text{Re}\Sigma^R&\simeq -\varepsilon\sqrt{\frac{\Lambda}{\pi |\varepsilon| }},
\\
\Gamma&\simeq \sqrt{\frac{|\varepsilon|\Lambda}{\pi}},
\label{critGamma}
\end{align}
which is in agreement with the large-$N$ mean-field result of Refs. \cite{PhysRevB.33.3257,PhysRevB.33.3263}, and \cite{PhysRevB.89.014205}.
The subleading (for $\varepsilon/\Lambda\ll 1$) corrections to $\text{Re}\Sigma^R$ and $\Gamma$ 
are linear in $\varepsilon$ and $|\varepsilon|$, respectively, see Appendix \ref{app:self-energy}.
Equations (\ref{crit}) and (\ref{critGamma}) imply that the dynamical critical exponent $z$ within SCBA is $z=2$. 
The result for the critical regime is valid under the condition that is opposite to that in Eq.~(\ref{DOS-weak}),
\begin{equation}
|\beta-1|\ll \sqrt{2\pi|\varepsilon|/\Lambda}.
\label{border}
\end{equation}
If the  disorder is slightly away from the critical value $\beta=1$ [but the system is still 
in the critical regime (\ref{border})], the leading behavior (\ref{critGamma}) acquires a correction
$\delta\Gamma \simeq \Lambda (\beta-1)/\pi$.
The condition (\ref{border}) determines the product of the correlation-length exponent $\nu$
and the dynamical exponent $z$ within the SCBA:  $\nu z=2$. In combination with $z=2$ it yields $\nu = 1$.
As follows from Eq.~(\ref{critGamma}), the critical density of states scales as a square root of energy: 
\begin{equation}
\rho(\varepsilon)\simeq \frac{\Lambda^{3/2} |\varepsilon|^{1/2}}{\pi^{7/2}v^3}, \qquad \beta=1,
\label{DOS-crit}
\end{equation}
where the critical exponent for the scaling with energy is given by $d/z-1$ with $d=3$ and $z=2$.

Next, we discuss the energy dependence of the self-energy in the case of strong disorder $\beta>1$.  
 Outside of the critical regime, i.e., under the condition opposite to Eq.~(\ref{border}),  the imaginary part of the self-energy is mainly determined by the zero-energy result, Eq.~\eqref{strongG}, with energy-dependent corrections  proportional to $\varepsilon^2$. 
The real part is obtained by an expansion in low energies, as performed in Appendix~\ref{app:self-energy}, leading to
\begin{align}
\label{strong}
\text{Re}\Sigma^R\approx\frac{\tilde{\beta}-2}{\tilde{\beta}-1}\varepsilon,
\end{align}
where the renormalized dimensionless disorder strength is defined as 
\begin{align}
\tilde{\beta}=\beta \frac{\Lambda^2}{\Lambda^2+\Gamma^2(\varepsilon=0)}.
\label{tilde-beta}
\end{align}
In the limit of very strong disorder, using Eq.~(\ref{gamma0}),
we get
\begin{equation}
\tilde{\beta}\simeq \frac{3\beta}{\beta+3}\to 3, \qquad \beta\to \infty.
\label{tbeta-inf}
\end{equation}
This saturation of the renormalized disorder strength will be of key importance
for establishing the strong-disorder asymptotic behavior of the conductivity
in Sec. \ref{sec:SCBAcon} below.


\begin{figure}
\begin{center}
\includegraphics[width=7.5cm]{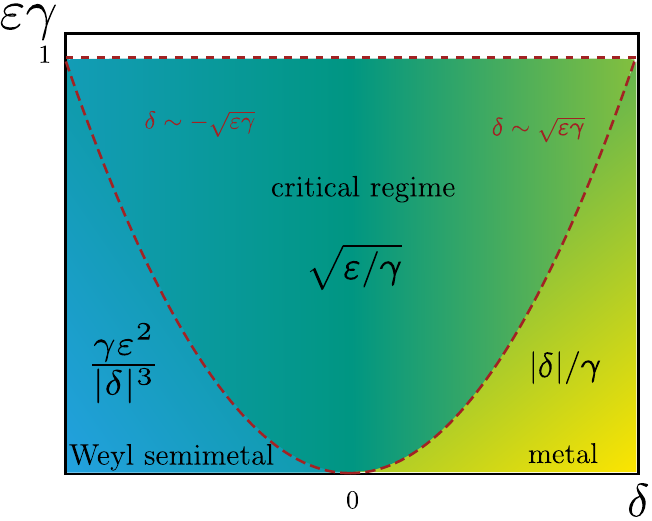} 
\caption{Scaling of the imaginary part of self-energy and the density of states with energy $\varepsilon$ in the three regimes (Weyl semimetal, critical region, diffusive metal) depending on the strength of disorder characterized by $\delta=1-\beta$ and on energy. The diagram clearly shows the regimes (``phases'') of weak (blue region), critical (green), and strong disorder (yellow). The borders of the regimes are indicated by red dashed lines. }
\label{DoS}
\end{center}
\end{figure}


The scaling of $\Gamma$ [and thus of density of states according to Eq.~(\ref{DOS-def})] in different regions of the parameter plane spanned by the disorder and the energy is presented in Fig.~\ref{DoS}.  This plot has an appearance characteristic for a vicinity of a quantum critical point: the critical regime separating the Weyl-semimetal and the metallic phases. It is seen that, for $\beta$ not far from the critical value $\beta=1$,  the system enters with increasing energy the critical regime with a square-root energy dependence of the density of states. 

\begin{figure}
\begin{center}
\includegraphics[width=7.5cm]{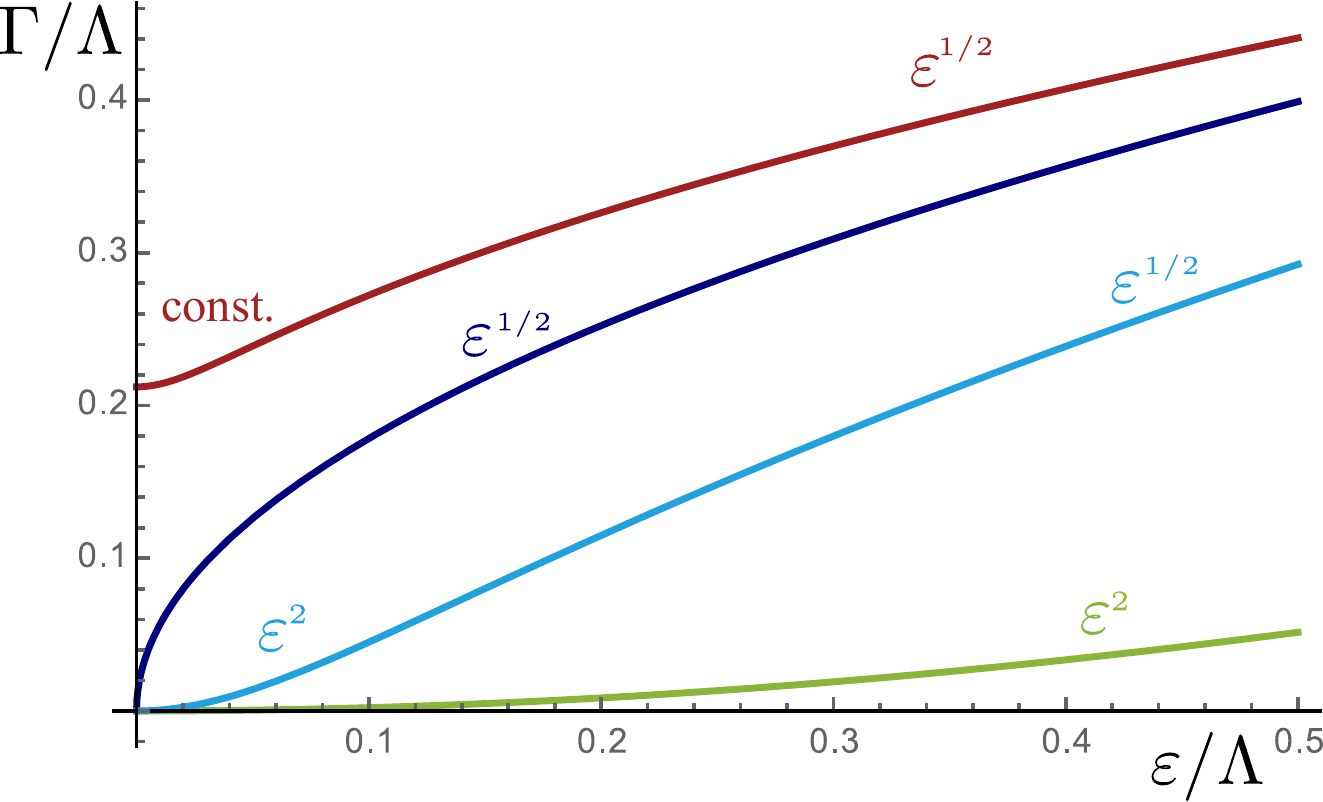} 
\caption{Imaginary part of the self-energy as a function 
of energy $\varepsilon$ obtained by numerically solving Eq.~\eqref{imag} and \eqref{real} for weak disorder, $\beta=0.1$ (green curve) and $\beta=0.5$ (light blue), critical disorder $\beta=1$ (dark blue), and strong disorder $\beta=2.3$ (red). 
The results illustrate analytical asymptotics given by Eqs.~(\ref{strongG}), (\ref{weak}), (\ref{critGamma}).
}
\label{fig:image}
\end{center}
\end{figure}


\begin{figure}
\begin{center}
\includegraphics[width=7.5cm]{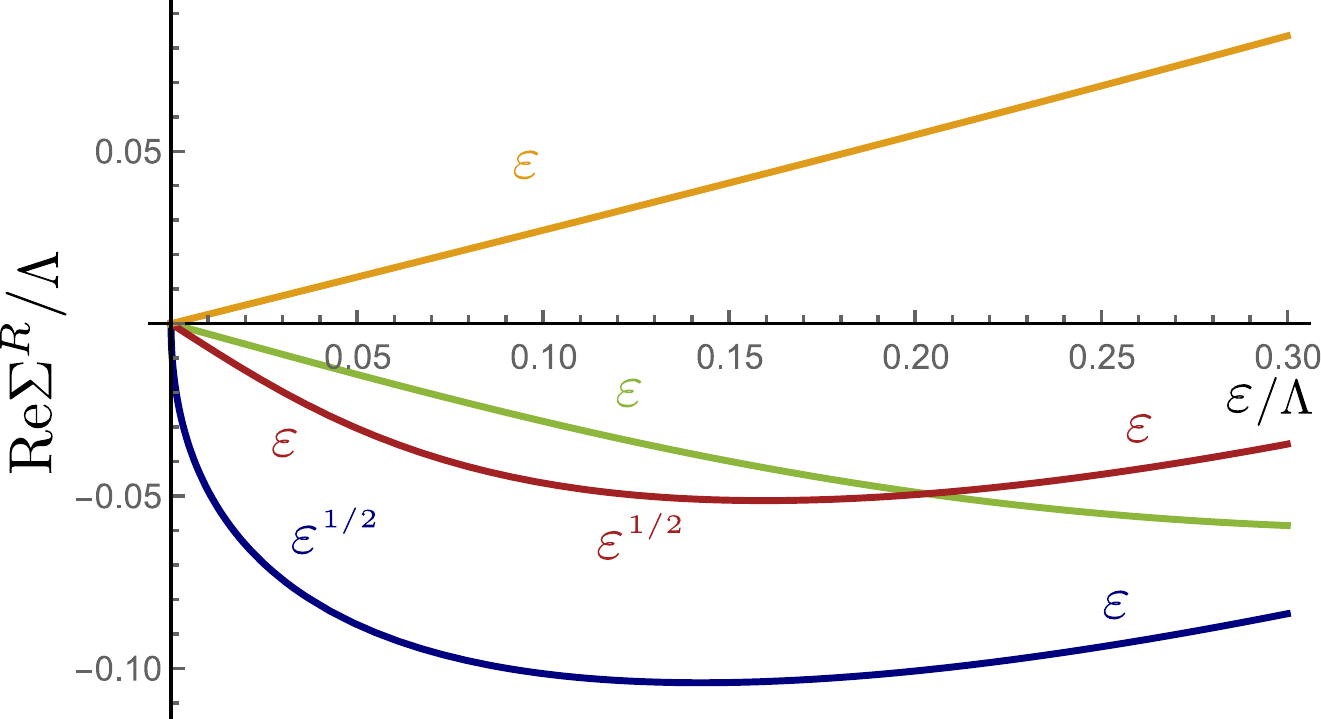} 
\caption{Real part of the self-energy as a function of energy $\varepsilon$ obtained by numerically solving Eqs.~\eqref{real} and \eqref{imag} for 
weak disorder, $\beta=0.23$ (green curve), critical disorder, $\beta=1$ (dark blue), and strong disorder, $\beta=1.8$ (red) and $\beta=5$ (orange).  The results illustrate analytical asymptotics given by Eqs.~(\ref{weak}), (\ref{crit}), and (\ref{strong}).
}
\label{fig:reale}
\end{center}
\end{figure}


Let us now compare the analytical results with the numerical evaluation of Eq.~\eqref{self-energy}. 
The imaginary part of the self-energy at two values of the bare energy $\varepsilon$ 
is shown, along with the $\varepsilon=0$ curve,  in Fig.~\ref{fig:imagpart}. The critical smearing of the transition is evident. To better visualize the $\varepsilon$ dependence of the imaginary part in different regimes, we show it as a function of energy at various $\beta$ in Fig.~\ref{fig:image}. All three types of behavior (semimetallic $\varepsilon^2$, critical $\varepsilon^{1/2}$, and metallic $\varepsilon^0$) are perfectly observed. In particular, Fig.~\ref{fig:image} illustrates the crossover from either semi-metallic or metalic behavior to the critical regime with increasing energy, as implied by the ``phase diagram'', Fig.~\ref{DoS}.

\begin{figure}
\begin{center}
\includegraphics[width=8cm]{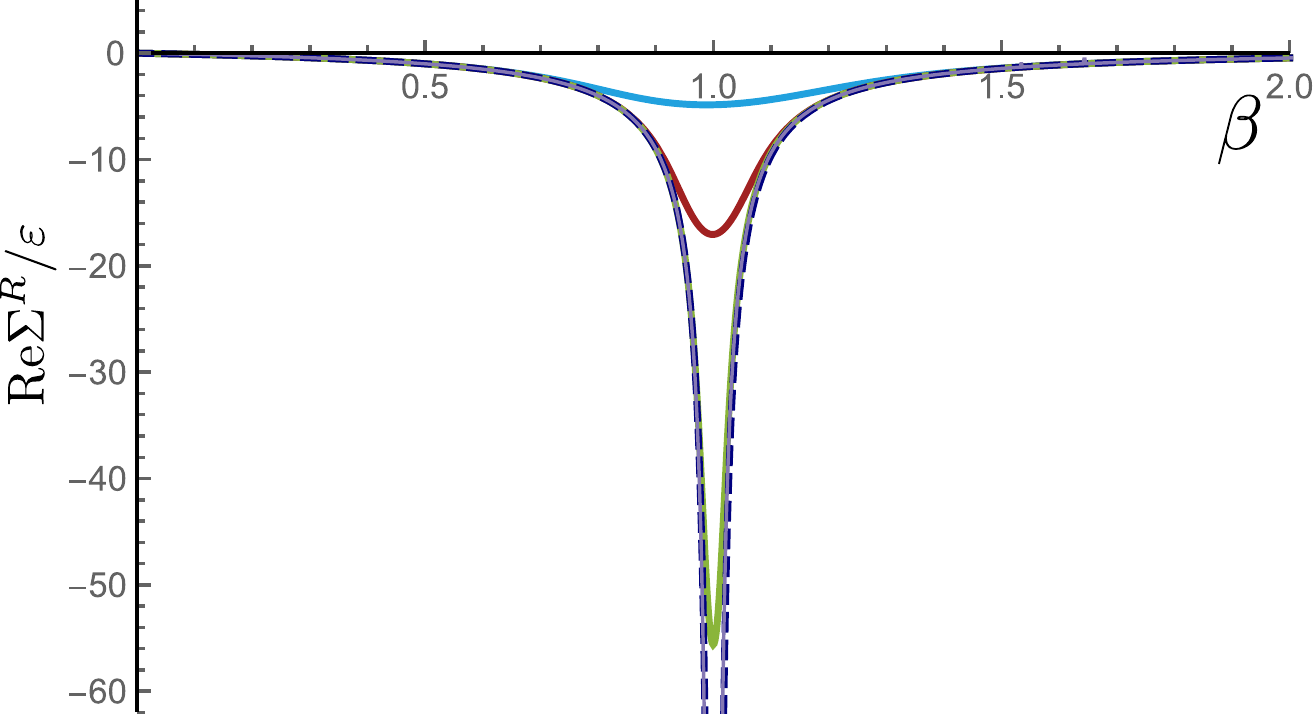} 
\caption{Real part of the self-energy (divided by energy $\varepsilon$) obtained from the numerical solution of Eq.~\eqref{real} as a function of 
$\beta$ for different values of $\varepsilon$. The green, red, and blue curves correspond to $\varepsilon/\Lambda=10^{-4}, 10^{-3}, 10^{-2}$, respectively.  The dark-blue dashed curve represents the limit $\varepsilon\to 0$. The results illustrate analytical asymptotics given by Eqs.~(\ref{weak}), (\ref{crit}), (\ref{strong}).
}
\label{fig:realpart}
\end{center}
\end{figure}

\begin{figure}
\begin{center}
\includegraphics[width=7.5cm]{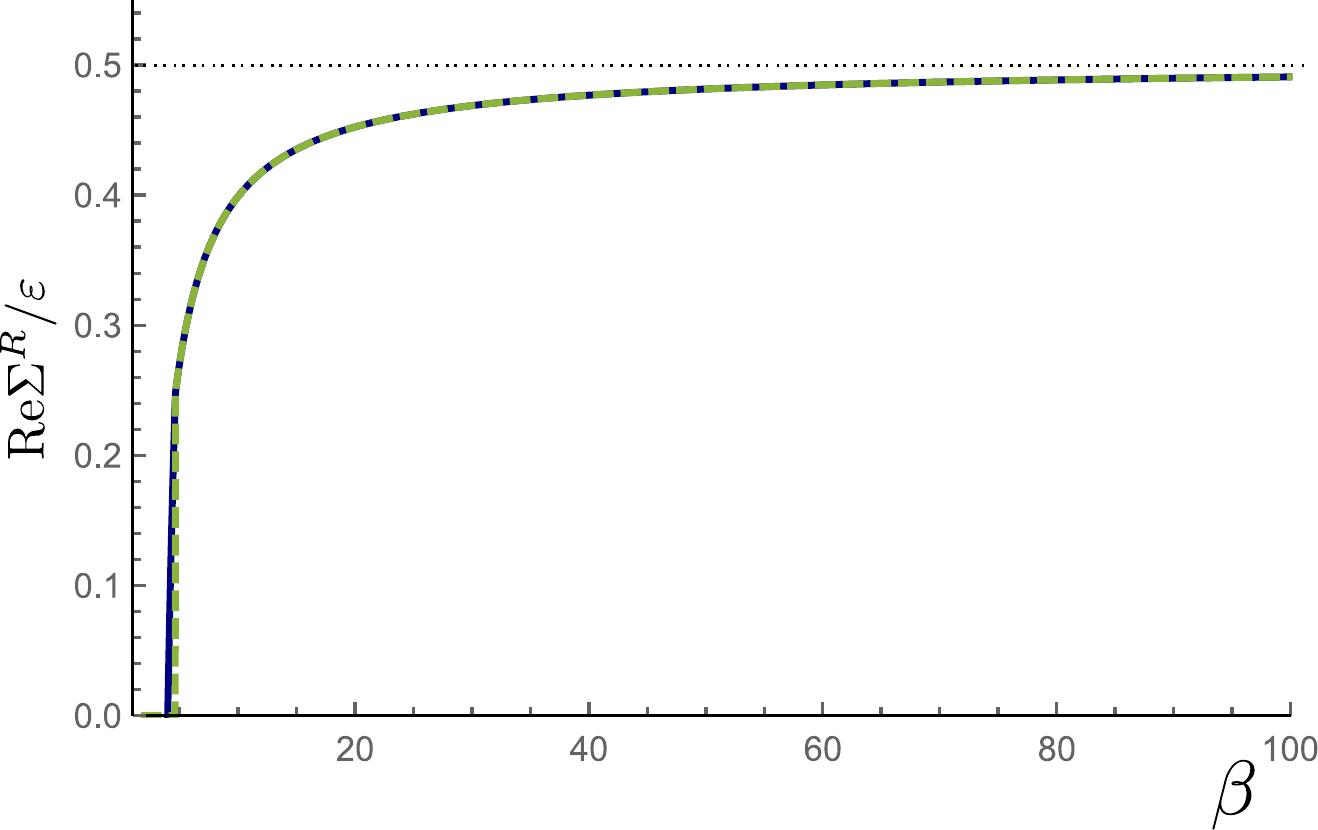} 
\caption{Real part of the self-energy (divided by $\varepsilon$) as a function of $\beta$ for $\beta \gg 1$. Green dashed line: numerical solution of Eq.~\eqref{real}.  The black solid curve corresponds to the solution of Eq.~\eqref{realpart} with the numerical solution for $\Gamma$ given by Eq.~\eqref{imag}. 
}
\label{fig:reallarge}
\end{center}
\end{figure}

Figures \ref{fig:realpart} and \ref{fig:reale} illustrate the behavior of the real part of the self-energy. In agreement with Eqs.~(\ref{weak}) and
(\ref{strong}), the real part scales linearly in $\varepsilon$ both in semimetalic and metallic regime. The corresponding coefficient $\text{Re}\Sigma^R /\varepsilon$ is small far away from the critical point and diverges when $\beta$ approaches unity, see dashed line in Fig.~\ref{fig:realpart}. At fixed energy, the divergence is avoided by the crossover to the critical regime.  In the critical regime (\ref{border}) the real part scales as $\varepsilon^{1/2}$ as predicted by Eq.~(\ref{crit}). 
Figure \ref{fig:reallarge} illustrates the behavior at very strong disorder, $\beta\gg 1$, where $\text{Re}\Sigma^R/\varepsilon\to 1/2$. This behavior will be important for the analysis of the conductivity in the next section. 

It is worth mentioning that, compared to Ref.~\cite{2018arXiv180809860S} where the density of states was analyzed as a function of the disorder strength only at $\varepsilon=0$, our results describe also the energy-dependence of the density of states. A related advantage of our SCBA analysis is its essentially analytical character, which should be contrasted to the computational self-consistent approximation (``Schwinger-Dyson-Ward approximation'') of Ref.~\cite{2018arXiv180809860S}. 

It is also instructive to compare the SCBA results for the self-energy with those obtained by one-loop RG approach.
The energy renormalization at criticality considered in Ref.~\cite{PhysRevB.91.035133} translates to
\begin{equation}
\varepsilon-\text{Re}\Sigma^R(\varepsilon) \sim \varepsilon \sqrt{\frac{\Lambda}{v K(\varepsilon)}},
\label{RG-real}
\end{equation}
where $K(\varepsilon)$ is the energy dependent momentum scale that satisfies the self-consistent 
condition of the RG flow termination,
\begin{equation}
\varepsilon-\text{Re}\Sigma^R(\varepsilon) \sim v K(\varepsilon).
\label{RG-termin}
\end{equation}  
It follows from Eqs.~(\ref{RG-real}) and (\ref{RG-termin}) that 
\begin{equation}
v K(\varepsilon)\sim \varepsilon^{2/3}\Lambda^{1/3}, \qquad 
\text{Re}\Sigma^R(\varepsilon) \sim \varepsilon 
\left(1-\frac{\Lambda^{1/3}}{\varepsilon^{1/3}}\right).
\label{RG-result}
\end{equation}
The difference between Eq.~(\ref{RG-result}) and the SCBA result (\ref{crit}) is that the square-root renormalization factor in the RG calculation is cut off by $v K(\varepsilon)$ rather than by $\varepsilon$. As a result, the dynamical exponents differ in the one-loop RG and SCBA approaches: 
$
z=3/2$ vs. 
$
z=2$, respectively.

\section{Conductivity within SCBA}
\label{sec:SCBAcon}

We calculate now the conductivity $\sigma_{xx}$ of a Weyl semimetal for weak, strong and critical disorder within
the pointlike disorder model discussed above.
We use the Kubo formula for the real part of the conductivity, reading
\begin{align}\label{Kubo}
&\sigma_{xx}(\omega,T)
=
\text{Re}\int\frac{d\varepsilon}{2\pi}\frac{f_{T}(\varepsilon)}{\omega}\int\frac{d^{3}\textbf{p}}{(2\pi)^{3}}
\nonumber\\
&\quad \times 
\text{Tr}
\left\{\left[\hat{G}^{R}(\varepsilon,\textbf{p})-\hat{G}^{A}(\varepsilon,\textbf{p})\right]
\hat{j}_{x}^\text{tr}
\hat{G}^{A}(\varepsilon-\omega,\textbf{p})\hat{j}_{x}
\right.\nonumber\\
&\quad \left.
+\hat{G}^{R}(\varepsilon+\omega,\textbf{p})\hat{j}_{x}^\text{tr}\left[\hat{G}^{R}(\varepsilon,\textbf{p})
-\hat{G}^{A}(\varepsilon,\textbf{p})\right]\hat{j}_{x}\right\}.
\end{align}
Here $\hat{j}_x=ev\sigma_x$ is the bare current operator and $\hat{j}_x^\text{tr}$
is the current vertex dressed by disorder and dependent on the external frequency $\omega$. The dressed vertex is discussed in Appendix~\ref{app:vertex}. 

The importance of vertex corrections in Weyl semimetals in the dc limit and for weak disorder was discussed in Ref.~\cite{PhysRevB.89.014205}. Here, we consider the effect of vertex correction also for the ac conductivity and in the full range of disorder. We find that the vertex corrections are of particular importance in the regimes of critical and strong disorder. 

\begin{widetext}

After performing the momentum integration, the conductivity reads
\begin{align}\label{calccon}
\sigma_{xx}(\omega, T)=&\frac{e^2v^2}{3\gamma}\int_{-\infty}^{\infty}\frac{d\varepsilon}{2\pi}\frac{f_T(\varepsilon)-f_T(\varepsilon +\omega)}{\omega}\nonumber\\
\times\, &\text{Re}\left\{
\frac{v}{v-v_{x}^{RA}(\varepsilon,\omega)}\left[\frac{\Sigma^R(\varepsilon+\omega)-\Sigma^A(\varepsilon)}{\Sigma^R(\varepsilon+\omega)-\Sigma^A(\varepsilon)-\omega}+\frac{\Sigma^R(\varepsilon+\omega)+\Sigma^A(\varepsilon)}{2\varepsilon+\omega-\Sigma^R(\varepsilon+\omega)-\Sigma^A(\varepsilon)}\right]
\right.\nonumber\\
&\left.\quad
-\frac{v}{v-v_{x}^{RR}(\varepsilon,\omega)}\left[\frac{\Sigma^R(\varepsilon+\omega)-\Sigma^R(\varepsilon)}{\Sigma^R(\varepsilon+\omega)-\Sigma^R(\varepsilon)-\omega}+\frac{\Sigma^R(\varepsilon+\omega)+\Sigma^R(\varepsilon)}{2\varepsilon+\omega-\Sigma^R(\varepsilon+\omega)-\Sigma^R(\varepsilon)}\right]
\right\},
\end{align}
where $v_{x}^\text{RA/RR}(\varepsilon,\omega)$ are calculated in Appendix~\ref{app:vertex}. Using  Eqs.~(\ref{vertRR}) and (\ref{vertRA}), we express the conductivity in terms of the self-energies as
\begin{align}\label{cond-explicit}
\sigma_{xx}(\omega, T)=
&\frac{2e^2v^2}{\gamma}\int_{-\infty}^{\infty}\frac{d\varepsilon}{2\pi}\frac{f_T(\varepsilon)-f_T(\varepsilon +\omega)}{\omega}\nonumber\\
\times\,&\text{Re}\left\{
\frac{(\varepsilon+\omega)\Sigma^A(\varepsilon)-\varepsilon \Sigma^R(\varepsilon+\omega)}
{\left[\varepsilon-\Sigma^A(\varepsilon)\right]\left[3\omega+4\Sigma^A(\varepsilon)\right] 
+\left[\varepsilon+\omega-\Sigma^R(\varepsilon+\omega)\right]\left[3\omega-4\Sigma^R(\varepsilon+\omega)\right]}
- \left[ \Sigma^A(\varepsilon)\to \Sigma^R(\varepsilon)\right]
\right\}.
\end{align}  
\end{widetext}

\subsection{Weak and critical disorder}

In the regime of weak disorder, using Eq.~\eqref{weak} for the self-energy, 
we calculate the conductivity in the two most interesting limits $T=0$ and $\omega=0$. 
We start with the case of $\omega=0$. The conductivity is dominated by
the retarded-advanced contribution in Eq.~(\ref{calccon})
\begin{align}
\sigma_{xx}(T,\omega=0)&\simeq\frac{e^2v^2}{2\pi\gamma}
\frac{1-\beta}{1+\beta}=\frac{e^2\Lambda}{4\pi^3v}\frac{1-\beta}{\beta(1+\beta)},
\label{sigma-T-weak}
\end{align}
which does not depend on temperature.
Only at the crossover to the regime of critical disorder at $T\sim (1-\beta)^2 \Lambda$ does the 
$T$-dependent retarded-retarded term 
$$\sigma_{xx}^{RR}(T,\omega=0)\sim \frac{e^2 T^2}{v\Lambda}\frac{\beta}{(3+\beta)(1-\beta)^3}$$
become comparable to the main contribution.
We emphasize that the condition of validity of Eq.~(\ref{sigma-T-weak}) is ${T\ll\Lambda(1-b)^2}$
which agrees with the condition (\ref{border}) for the border separating the regimes of weak and critical disorder. This means that in the limit $T\to 0$ Eq.~(\ref{sigma-T-weak}) is valid for all $\beta<1$.
 
For finite $\omega$ and $T=0$, the integral over $\varepsilon$ is dominated by the 
point $\varepsilon=-\omega/2$ in the retarded-retarded contribution in Eq.~(\ref{calccon}). 
Evaluating the integral around this point, we get a linear frequency dependence of the 
conductivity with the disorder-dependent coefficient:
\begin{align}\label{calccon1}
\sigma_{xx}(T=0,\omega)
         &\simeq \frac{e^2 |\omega|}{4 \pi v (1 - \beta) (3 + \beta)}.         
\end{align}
Similarly to Eq.~(\ref{sigma-T-weak}), the condition of validity of this expression 
is $\omega\ll \Lambda (1-b)^2$, which again means that in the limit $\omega\to 0$ 
 the range of the applicability of the weak-disorder formula (\ref{calccon1}) extends up to $\beta\to 1$.
For small $\beta\to 0$, we obtain
\begin{align}
\sigma_{xx}(T=0,\omega)&=\frac{e^2 |\omega|}{12 \pi v}\left(1+\frac{2}{3}\beta\right),
\end{align}
which agrees with the results of Ref.~\cite{PhysRevLett.108.046602} and Ref.~\cite{Roy}.
We see that the limits of $\omega=0$ and $T=0$ are not interchangeable, as discussed in Ref.~\cite{PhysRevLett.108.046602}. Furthermore, we find that the dc conductivity vanishes at the critical-disorder point. We would like to stress that the vanishing of the ac conductivity at $\omega\to 0$ is related to the vanishing density of states in the regime of weak disorder within the SCBA scheme. 

For the calculation of the conductivity at critical disorder, we use Eqs.~\eqref{crit} and \eqref{critGamma}
for the self-energies.
For $\omega=0$, the result is
\begin{align}
\sigma_{xx}(T,\omega=0)=\mathcal{C}_T\frac{e^2}{2\pi v}
\sqrt{\Lambda T}.
\label{crit-T}
\end{align}
The numerical prefactor for the hard-cutoff model used in this paper is given by
$\mathcal{C}_T=3(1-\sqrt{2})\zeta(1/2)/4\pi\approx 0.14$,
where $\zeta(x)$ is the Riemann zeta-function.
We see that the conductivity in the dc limit (finite $T$ and $\omega=0$) matches the result for weak disorder at $(1-\beta)^2\Lambda \sim T$. 
The calculation of the ac conductivity at $T=0$ and $\beta=1$ yields
\begin{align}
\sigma_{xx}(T=0,\omega)&=\mathcal{C}_\omega\frac{e^2}{2\pi v}\sqrt{\Lambda |\omega|},
\label{crit-omega}
\end{align}
where 
$$\mathcal{C}_\omega=\frac{1}{6\pi^{3/2}}\left[ \frac{29}{15}-\frac{\ln(3+2\sqrt{2})}{2\sqrt{2}}\right]\approx 0.04.$$
Expression (\ref{crit-omega}) matches Eq.~(\ref{calccon1}) at $\omega\sim \Lambda(1-\beta)^2$ and Eq.~(\ref{crit-T}) at $\omega\sim T$.
The obtained quantum critical scaling of the ac conductivity is in agreement with the general scaling form discussed in Ref.~\cite{Roy}. 

\subsection{Strong disorder}

Let us now discuss the regime of strong disorder $\beta>1$. We substitute the result for the self-energy for strong disorder, Eq.~\eqref{strongG}, into Eq.~(\ref{calccon}) and write Eq.~(\ref{strong}) with Eq.~(\ref{tilde-beta}) as
\begin{equation}
\text{Re}\Sigma^R=A\varepsilon, \qquad A=\frac{\tilde{\beta}-2}{\tilde{\beta}-1}.
\label{A-def}
\end{equation}
For $\omega\to 0$ and $T\to0$, the vertex corrections \eqref{vertRR} and \eqref{vertRA} simplify for strong disorder to
\begin{align}
v_{x}^{RR}&=-v\frac{2-A}{3(1-A)},\\
v_{x}^{RA}&=v\frac{A+1}{3(1-A)}.
\label{VC}
\end{align}
Vertex corrections need to fulfil the condition $v_{x}/v<1$. This is indeed the case for $A<1/2$ which is valid up to highest disorder strength $\beta\to\infty$, see Eq. (\ref{one-half}) in Appendix 
\ref{app:self-energy} and Fig.~\ref{fig:reallarge}. 
For lowest temperatures and frequencies, the conductivity reads
\begin{align}
\sigma_{xx}(T\to 0,\omega\to 0)
&=\frac{e^2v^2}{2\pi\gamma} \frac{(7-8A)}{(1-2A)(5-4A)}.
\label{cond-strong-disorder}
\end{align}
We note that, in contrast to the weak-disorder regime, Eq. (\ref{sigma-T-weak}), the RR contribution is not small compared to the RA one:
$$
\sigma_{xx}^{RA}=\frac{e^2v^2}{2\pi\gamma(1-2A)},
\qquad \sigma_{xx}^{RR}=\frac{e^2v^2}{\pi\gamma(5-4A)}.
$$

We find that the condition $A<1/2$ for the vertex corrections is manifested again in the calculation of the conductivity, where 
a positive conductivity without any singularities is obtained under this restriction. 
It is important to emphasize that the conductivity saturates as a function of disorder in the limit 
$\beta\gg 1$. Indeed, using Eqs.~(\ref{A-def}) and (\ref{one-half}), we see that 
$1-2A\to 9/5\beta$ for $\beta\to \infty$, which cancels the factor $\gamma\propto \beta$ in the denominator
of Eq.~(\ref{cond-strong-disorder}). The saturation value of the conductivity in the limit of strong disorder is then given by
\begin{align}
\sigma_{xx}
&\simeq \frac{5 e^2 \Lambda}{36 \pi^3 v}, \qquad \beta\gg 1.
\label{satur}
\end{align}

\begin{figure}
\begin{center}
\includegraphics[width=7.5cm]{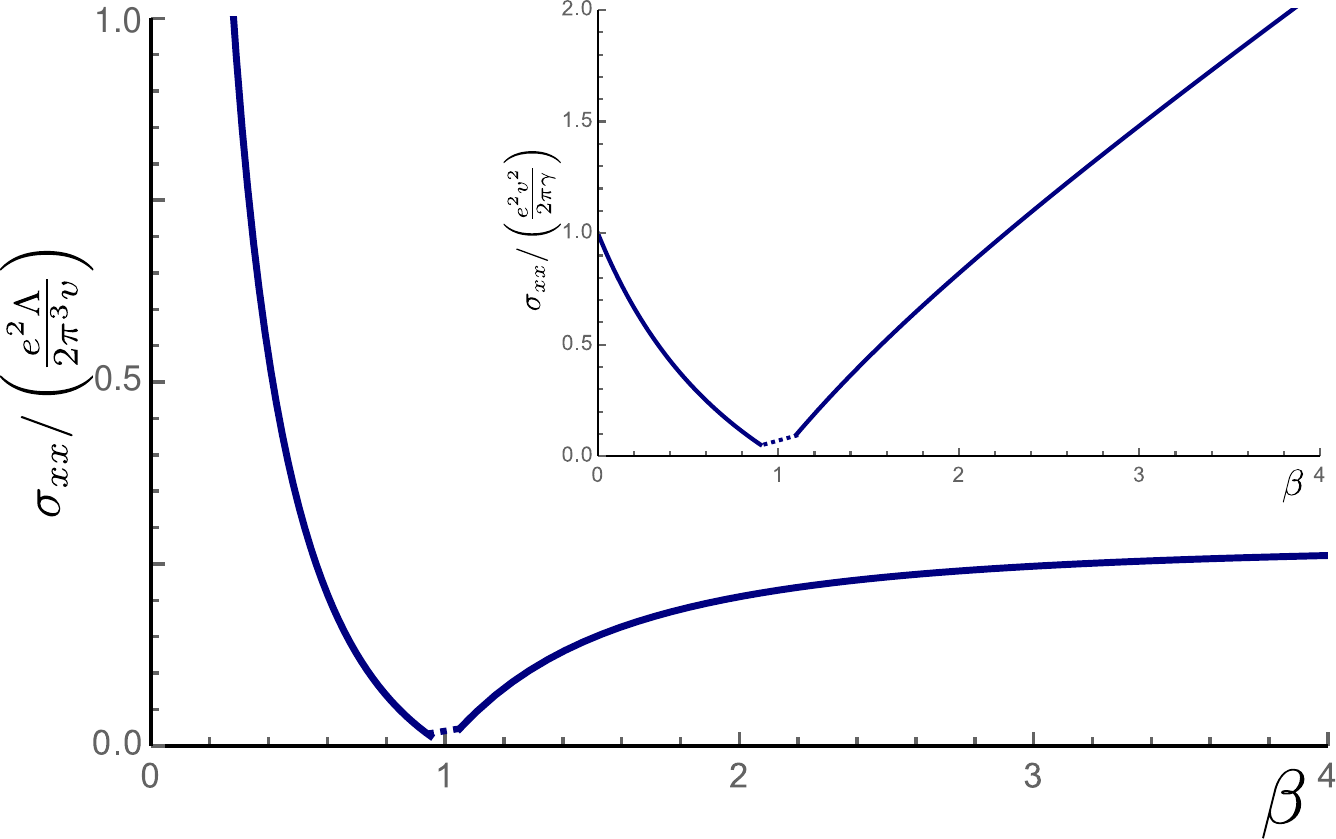}
\caption{Conductivity in the limit $\omega\to0$ and then $T\to0$ with the self-energies (real and imaginary part) obtained numerically from Eqs.~(\ref{real}) and (\ref{imag}). The dotted part corresponds to the region close to the critical point $\beta_c$ where the straightforward numerical evaluation is complicated by the divergence of $\text{Re}\Sigma^R$ for $\beta\to 1$ [in this region, the conductivity vanishes linearly with  $|\beta-1|$, see Eqs. (\ref{sigma-T-weak}) and (\ref{cond-beta-1})]. The conductivity saturates at the value $5/18\simeq 0.28$ in units of $e^2 \Lambda/(2\pi^3 v^2)$, see Eq.~(\ref{satur}). The inset depicts the conductivity in units of $e^2v^2/2\pi \gamma$ used for the conductivity at weak disorder (where $\text{Re}\Sigma^R\ll \varepsilon$), see Eq.~(\ref{sigma-T-weak}) at $\beta\to 0$. Thus, the inset emphasizes the important role of the real part of the self-energy and renormalization of the disorder strength, $\beta\to\tilde{\beta}$, in the vertex corrections for strong disorder.} 
\label{fig:conductivity}
\end{center}
\end{figure}

Furthermore, the conductivity vanishes at the critical disorder $\beta = 1$, where 
$A \simeq -1/(\beta-1)\to-\infty$:
\begin{align}
\sigma_{xx}&\simeq\frac{e^2v^2}{2\pi\gamma}(\beta-1), \qquad \beta\to 1.
\label{cond-beta-1}
\end{align} 
Note the factor of 2 compared to Eq. (\ref{sigma-T-weak}) in the dependence of the conductivity on 
$|\beta-1|$ around the critical point: this asymmetry is due to the RR contribution at $\beta>1$. 
The renormalization (\ref{tilde-beta}) of the dimensionless disorder strength $\beta$ can be neglected around $\beta=1$, but gets crucial for stronger disorder, ensuring that $A<1/2$. Thus, by fully incorporating the vertex corrections, one has to consistently keep track of the modification of the real part of the self-energy in the SCBA analysis for strong disorder. 

The dependence of the low-$T$, zero-frequency conductivity on the disorder strength $\beta$ as obtained by the numerical evaluation of the Kubo formula is demonstrated in Fig.~\ref{fig:conductivity}. The observed behavior confirms the analytical asymptotics (\ref{sigma-T-weak}),  (\ref{crit-T}), and (\ref{cond-strong-disorder}). 

The saturation of the conductivity at strong disorder should be contrasted with the result of Ref.~\cite{PhysRevB.89.054202}, where the conductivity was found to decrease as $1/\beta$. The  reason for this behavior is that in Ref.~\cite{PhysRevB.89.054202} the formula for vertex corrections derived for weak disorder has been used in the strong-disorder limit.

\subsection{Smooth disorder}

Considering the limit of large $\beta$ for pointlike impurities on a lattice model corresponds to a large potential on each lattice site which would completely destroy the model. Below, we consider a model of smooth disorder potential, where the limit of large $\beta$ is realized by increasing the correlation length instead of the amplitude of the potential. 

In analogy with the model of pointlike impurities considered above, we assume that the disorder potential is diagonal in both spin and pseudospin indices and neglect the internode scattering.
This impurity correlator relates the disorder strength $\gamma_b$ to the characteristic magnitude of disorder potential $U_0$ and its correlation radius $b$ as
\begin{align}\label{gamma_smooth}
\gamma_b=n_\text{imp} (U_0 b^3)^2.
\end{align}

The self-energy for smooth disorder is momentum-dependent and the SCBA requires a solution of coupled integral equations. 
Since the disorder correlator introduces a natural momentum cutoff replacing $\Lambda/v$, 
the results obtained for the pointlike disorder can be used to qualitatively describe the smooth-disorder case (cf., e.g., Refs.~\cite{PhysRevB.89.054202} and \cite{Klier}), 
with the replacements $\beta\to \beta_b$ and $\Lambda\to \Lambda_b$,
where 
$\beta_b$ and $\Lambda_b$ are given by
\begin{align}
\beta_b=\frac{\gamma_b}{2\pi^2v^2b}, \quad \Lambda_b=\frac{v}{b}. 
\end{align}
To show that large values of $\beta_b$ can be realized for relatively low impurity potential $U_0<\Lambda$, we rewrite the dimensionless disorder strength in terms of the bandwidth $\Lambda$ and the lattice constant $a=v/\Lambda$, assuming that the distance between the impurities is of the order of their correlation radius,
$n_\text{imp}\sim 1/b^3$:
\begin{align}
\beta_b\sim\left(\dfrac{U_0}{\Lambda}\right)^2\left(\dfrac{b}{a}\right)^2. 
\end{align}
This shows that large $\beta_b$ can be achieved for large $b\gg a$ even for small impurity potentials, $U_0\ll \Lambda$. 

The qualitative behavior of the density of states does not fundamentally change within the model of smooth disorder as compared that of point-like disorder. In particular, the density of states remains vanishing for $\beta_b<1$ for $\varepsilon=0$ and becomes finite above. In the limit of strong disorder, the broadening can be approximated by 
\begin{equation}
\Gamma_\text{smooth}\sim\Lambda_b\sqrt{\beta_b},
\end{equation} 
again in a full analogy with the case of pointlike impurities. 
This renders our results (\ref{cond-strong-disorder}) for the conductivity in the strong-disorder regime applicable to the model of smooth disorder, with $A$ defined by Eqs.~(\ref{A-def}), (\ref{strong}), and (\ref{tilde-beta}), where we replace $\beta\to\beta_b$, $\Lambda\to \Lambda_b$ and $\Gamma\to \Gamma_\text{smooth}$.

\section{Summary}
\label{sec:summary}

We considered the density of states and the conductivity of a Weyl semimetal within the SCBA in the full range of disorder strength, from weak ($\beta\ll 1$) to strong ($\beta\gg 1$) disorder, see Fig. \ref{DoS}. The limit of large $\beta$ can be realized in a smooth disorder model for rather weak impurity potentials.   
The density of states for weak disorder vanishes as $\varepsilon^2$, while the density of states for strong disorder is finite. For the regime of critical disorder, we find a density of states proportional to 
the square root of energy $\varepsilon$.

The conductivity for weak disorder is constant in the dc limit (first taking $\omega\to0$ and then $T\to0$). In the opposite limit, the weak-disorder ac conductivity is linear in $\omega$. In both limits, we derived the explicit dependence of the conductivity on $\beta<1$.
The conductivity at critical disorder, $\beta=1$, is proportional to $\sqrt{\omega}$ or $\sqrt{T}$, whichever is larger.

For the strong disorder, the renormalization of the dimensionless disorder strength $\beta$ ensures that the vertex corrections remain small $v_{x}/v<1$, leading to a saturation of the conductivity in the limit $\beta\to\infty$. This limit of very strong disorder with the saturating conductivity is realized within a model of smooth disorder, where the strong-disorder limit ($\beta_b\gg 1$) can be established by the large correlation length instead of a large magnitude of the impurity potential. For smooth disorder, the appearance of the critical point persists at the Weyl point. The SCBA density of states vanishes below 
$\beta_b\sim1$ and can be approximated with $\Gamma_\text{smooth}\sim\Lambda_b\sqrt{\beta_b}$ for strong disorder, thus leading to the saturation of the conductivity, Fig.~\ref{fig:conductivity}.

\vspace{0.5cm}

\acknowledgments

We acknowledge useful discussions with P. Ostrovsky and B. Sbierski.
The work was supported by Carl-Zeiss-Stiftung (J.K.) and by the Priority
Programme 1666 ``Topological Insulators'' of the Deutsche
Forschungsgemeinschaft (DFG-SPP 1666). 

\appendix
\begin{widetext}
\section{Details of the calculation of self-energy}
\label{app:self-energy}

In this appendix, we present details of the SCBA calculation of the Green function in the model with a point-like disorder. The self-consistent equation (\ref{self-energy}) for the self-energy at arbitrary energy $\varepsilon$, as obtained after momentum integration performed in the main text, Eq.~\eqref{selfedef}, reads
\begin{align}\label{fullselfe}
\varepsilon-\mathcal{E}-i \Gamma&=\beta(\mathcal{E}+i\Gamma)
\left[-1+\frac{(\mathcal{E}+i\Gamma)}{2\Lambda}
\ln\left(\frac{\mathcal{E}+i\Gamma+\Lambda}
{\mathcal{E}+i\Gamma-\Lambda}\right)\right],
\end{align}
where we introduced for brevity
\begin{equation}
\mathcal{E}\equiv \varepsilon-\text{Re}\Sigma^R, \quad \Gamma \equiv -\text{Im}\Sigma^R.
\label{calE}
\end{equation}
We are interested in the situation when the energy is much smaller than the cutoff scale, 
$\Lambda\gg |\mathcal{E}|$.
At the same time, the relation between the cutoff $\Lambda$ and the broadening $\Gamma$ can be arbitrary:
for weak and critical disorder, we will have $\Gamma\ll \Lambda$, 
whereas for strong disorder, we will find $\Gamma\gg \Lambda$, see Eq.~(\ref{gamma0}). 

We first consider the case $\Gamma\ll \Lambda$ and replace the logarithmic term in 
Eq.~(\ref{fullselfe}) by a constant $-i\pi$. The next term in the expansion of the logarithm at $\Lambda\to \infty$ is given by $2(\mathcal{E}+i \Gamma)/\Lambda$ and can be omitted for establishing the leading behavior of the self-energy.
This yields a quadratic equation for the complex quantity 
$\mathcal{E}+i \Gamma$:
\begin{align}\label{self-energy-pi}
\varepsilon&\simeq (1-\beta)(\mathcal{E}+i \Gamma)-i\pi\beta \frac{(\mathcal{E}+i\Gamma)^2}{2\Lambda},
\end{align}
whose solution is given by
\begin{align}
\frac{\mathcal{E}+i \Gamma}{\Lambda}=-i\frac{1-\beta}{\pi\beta} 
+ i\sqrt{\left(\frac{1-\beta}{\pi\beta}\right)^2-\frac{2i\varepsilon}{\pi \beta\Lambda}}.
\label{quadrat}
\end{align}
The sign in front of the square root is dictated by the requirement $\Gamma\geq 0$.
Taking the real and imaginary parts of the square root on the right-hand side of Eq.~(\ref{quadrat}), we arrive at Eqs.~(\ref{ReSigma-sqrt}) 
and (\ref{ImSigma-sqrt}) of the main text.
The results (\ref{crit}) and (\ref{critGamma}) follow immediately from Eq.~(\ref{quadrat}) at $\beta=1$. 

Let us now analyze the self-energy in the limit $\varepsilon\to 0$ in the full range of disorder, including 
strong disorder, $\beta\gg 1$. For definiteness, we assume $\varepsilon\geq 0$. The consideration below allows us to extract the subleading corrections to Eq.~(\ref{quadrat}) at weak and critical disorder and to calculate the self-energy for strong disorder on equal footing.
Using 
\begin{align}
\text{Re}\left\{\ln\left(\frac{\mathcal{E}+i\Gamma+\Lambda}
{\mathcal{E}+i\Gamma-\Lambda}\right) \right\}
&=
\ln\left(\frac{\sqrt{\left[\Lambda^2-\mathcal{E}^2-\Gamma^2\right]^2+(2\Gamma\Lambda)^2}}
{(\Lambda-\mathcal{E})^2+\Gamma^2}\right)
\underbrace{\approx}_{\mathcal{E}/\sqrt{\Lambda^2+\Gamma^2} \ll 1}
\frac{2 \Lambda \mathcal{E}}{\Lambda^2 + \Gamma^2}
\left[1+\frac{(\Lambda^2-3\Gamma^2)\mathcal{E}^2}{3(\Lambda^2+\Gamma^2)^2}\right]
, \notag
\\
-\text{Im}\left\{\ln\left(\frac{\mathcal{E}+i\Gamma+\Lambda}
{\mathcal{E}+i\Gamma-\Lambda}\right) \right\}
&=
\frac{\pi}{2}+
\arctan\frac{\Lambda^2-\mathcal{E}^2-\Gamma^2}{2\Gamma\Lambda}
\underbrace{\approx}_{\mathcal{E}/\sqrt{\Lambda^2+\Gamma^2} \ll 1}
2\arctan\frac{\Lambda}{\Gamma} 
- \frac{2\Gamma \Lambda \mathcal{E}^2}{(\Lambda^2+\Gamma^2)^2},
\notag
\end{align}
we split the self-consistent equation (\ref{fullselfe}) into the equations corresponding to the real and imaginary parts of the self-energy:
\begin{align}
\varepsilon-\mathcal{E}&=-\beta \mathcal{E}
+\beta\frac{\mathcal{E}^2-\Gamma^2}{2\Lambda}
\ln\left(\frac{\sqrt{\left[\Lambda^2-\mathcal{E}^2-\Gamma^2\right]^2+(2\Gamma\Lambda)^2}}
{(\Lambda-\mathcal{E})^2+\Gamma^2}\right)
+\beta\,\frac{\mathcal{E}\Gamma}{\Lambda}
\left(\frac{\pi}{2}+
\arctan\frac{\Lambda^2-\mathcal{E}^2-\Gamma^2}{2\Gamma\Lambda}\right),
\label{real}
\\
\Gamma&=\beta\Gamma-\beta\frac{\mathcal{E}\Gamma}{\Lambda}
\ln\left(\frac{\sqrt{\left[\Lambda^2-\mathcal{E}^2-\Gamma^2\right]^2+(2\Gamma\Lambda)^2}}
{(\Lambda-\mathcal{E})^2+\Gamma^2}\right)
+\beta\,\frac{\mathcal{E}^2-\Gamma^2}{2\Lambda}
\left(\frac{\pi}{2}
+\arctan\frac{\Lambda^2-\mathcal{E}^2-\Gamma^2}{2\Gamma\Lambda}\right).
\label{imag}
\end{align} 

In the range of weak and critical disorder, we have $\Lambda\gg \Gamma$ and 
$\arctan(\Lambda/\Gamma)\simeq \pi/2-\Gamma/\Lambda$, which allows us to 
simplify Eqs.~(\ref{real}) and (\ref{imag}):
\begin{align}\label{re-weak-crit}
\varepsilon &\simeq (1-\beta)\mathcal{E}
+ \pi \beta\,\frac{\mathcal{E}\Gamma}{\Lambda} 
,\\
\label{im-weak-crit}
\Gamma&\simeq\beta\Gamma+ \pi \beta\,\frac{\mathcal{E}^2-\Gamma^2}{2\Lambda}
-2\beta \frac{\Gamma \mathcal{E}^2}{\Lambda^2}.
\end{align}
For weak disorder, we reproduce Eqs.~(\ref{weak}) from the main text, which were obtained there 
by expanding Eqs.~(\ref{ReSigma-sqrt}) 
and (\ref{ImSigma-sqrt}):
\begin{align}
\mathcal{E}\simeq \frac{\varepsilon}{1-\beta}, 
\qquad \Gamma\simeq \frac{\pi \beta \varepsilon^2}{2\Lambda(1-\beta)^3}.
\end{align}
For the critical disorder, $\beta=1$, 
we get
\begin{align}
\mathcal{E}\simeq \frac{\varepsilon \Lambda}{\pi \Gamma}, 
\qquad \Gamma^2\simeq \mathcal{E}^2-\frac{4}{\pi^2}\varepsilon \mathcal{E},
\end{align}
and 
refine the result of Eq.~(\ref{quadrat}) by including the leading corrections to $-i\pi$ in the logarithmic term in Eq.~(\ref{fullselfe}):
\begin{align}
\Gamma\simeq \sqrt{\frac{\varepsilon \Lambda}{\pi}} - \frac{\varepsilon}{\pi^2},
\qquad 
\mathcal{E} \simeq \sqrt{\frac{\varepsilon \Lambda}{\pi}} + \frac{\varepsilon}{\pi^2}.
\end{align} 
For an arbitrary sign of $\varepsilon$, this translates into 
\begin{align}
\Gamma\simeq \sqrt{\frac{|\varepsilon| \Lambda}{\pi}} - \frac{|\varepsilon|}{\pi^2},
\qquad
\text{Re}\Sigma^R(\varepsilon)\simeq -\varepsilon\left(\sqrt{\frac{\Lambda}{\pi|\varepsilon|}}+1-\frac{1}{\pi^2}\right),
\end{align}
where small corrections to Eqs.~(\ref{critGamma}) and (\ref{crit}) of the main text are included.

Let us now turn to the case of strong disorder. In this regime, 
$\Gamma$ is finite already at $\varepsilon=0$. 
In order to calculate the real part of self-energy, one can keep only the linear-in-$\mathcal{E}$ terms    
in Eq.~(\ref{real}), which also implies using there $\Gamma_0=\Gamma(\varepsilon=0)$ from Eq.~(\ref{gamma0}):
\begin{align}\label{strong-real}
&\varepsilon \simeq (1-\beta)\mathcal{E}
-\beta\mathcal{E}\,\frac{\Gamma^2_0}{\Lambda^2 + \Gamma^2_0}
+2\beta\mathcal{E}\,\frac{\Gamma_0 }{\Lambda}
\arctan\frac{\Lambda}{\Gamma_0},\\
\label{strong-im}
&\beta-1\simeq \beta\,\frac{\Gamma_0}{\Lambda}\arctan\frac{\Lambda}{\Gamma_0}.
\end{align}
This yields 
\begin{align}\label{realpart}
\text{Re}\Sigma=\varepsilon \left(1-\frac{1}{|\tilde{\beta}-1|}\right)
\end{align}
with the renormalization of the dimensionless disorder strength according to
\begin{align}
\label{tilde-beta-new}
\tilde{\beta}=\beta \frac{\Lambda^2}{\Lambda^2+\Gamma_0^2}.
\end{align}
Using the asymptotics for $\Gamma_0$ from Eq.~(\ref{strongG}), we write the real part of self-energy explicitly in terms of $\beta$: 
\begin{align}\label{strongRe}
\text{Re}\Sigma=
\varepsilon\times  \left\lbrace\begin{array}{ccc}
&-\dfrac{1}{(\beta-1)}, \quad &\beta-1\ll 1;
\\[0.3cm]
&\dfrac{1}{2}, \quad &\beta\gg 1.
\end{array}\right.
\end{align}
In the limit of strong disorder, including the correction to the second line of 
Eq.~(\ref{strongRe}), 
\begin{equation}
\Gamma_0\simeq \sqrt{\frac{\beta}{3}}\left(1-\frac{9}{10\beta}\right), \quad \beta\to \infty,
\end{equation}
we obtain 
\begin{equation}
\frac{\text{Re}\Sigma}{\varepsilon}\simeq \frac{1}{2}-\frac{9}{10\beta}, \quad \beta\to \infty.
\label{one-half}
\end{equation}

It is interesting to notice that Eqs. (\ref{realpart}) and (\ref{tilde-beta-new}) turn out to be applicable also for weak disorder, where $\Gamma_0=0$ and hence $\tilde\beta=\beta$, cf. Eq.~(\ref{weak}).  
Further, using Eq.~(\ref{realpart}), we calculate a small energy-dependent correction to $\Gamma_0$:
\begin{align}
\Gamma(\varepsilon)=\Gamma_0+\left[\tilde{\beta}
\frac{\Lambda^4+\Gamma_0^4}{\Lambda^2(\Lambda^2+\Gamma_0^2)}-1\right] \frac{\varepsilon^2}{\Gamma_0(\tilde{\beta}-1)^3}.
\end{align}
Here, the structure $\varepsilon^2/(\tilde{\beta}-1)^3$ is again reminiscent of Eq.~(\ref{weak}).
The above results for the self-energy are used in the main text to analyze the density of states and the conductivity in the full range of disorder strength.

\section{Frequency dependent vertex corrections}
\label{app:vertex}

This appendix is devoted to the evaluation of the vertex corrections in the Kubo formula for conductivity.  Assuming a finite external frequency $\omega$, the vertex corrections to the current vertex 
are given by the geometric series 
\begin{align}
\hat{G}^R(\varepsilon+\omega)\hat{j}^\text{tr}_x \hat{G}^{R}(\varepsilon)
&=\frac{v}{v-v_x^{RR}}\hat{G}^R(\varepsilon+\omega)\hat{j}_x \hat{G}^{R}(\varepsilon),\\
\hat{G}^R(\varepsilon+\omega)\hat{j}^\text{tr}_x \hat{G}^{A}(\varepsilon)
&=\frac{v}{v-v_x^{RA}}\hat{G}^R(\varepsilon+\omega)\hat{j}_x \hat{G}^{A}(\varepsilon),
\end{align}
where
\begin{align}
v_{x}^{\text{RR/RA}}(\varepsilon,\omega)&=\frac{v\gamma}{2}\int\frac{d^3p}{(2\pi)^3}
\text{Tr}\sigma_x\hat{G}^R(\varepsilon+\omega,\textbf{p})\sigma_x\hat{G}^{R/A}(\varepsilon,\textbf{p})\nonumber
\\
&=
v\gamma\int\frac{d^3p}{(2\pi)^3}
\frac{[\varepsilon+\omega -\Sigma^R(\varepsilon+\omega)][\varepsilon -\Sigma^{R/A}(\varepsilon)]-v^2 p^2_z}{\left\{[\varepsilon+\omega -\Sigma^R(\varepsilon+\omega)]^2-v^2p^2\right\}
\left\{[\varepsilon -\Sigma^{R/A}(\varepsilon)]^2-v^2p^2\right\}}.
\end{align}
Evaluation of the momentum integrals, using the self-consistency equation (\ref{selfedef}), yields:
\begin{align}
v_{x}^{\text{RR/RA}}(\varepsilon,\omega)&=\frac{v}{3}
\left[-1+\frac{\omega}{\Sigma^{R}(\varepsilon+\omega)-\Sigma^{R/A}(\varepsilon)-\omega}
+\frac{2(2\epsilon+\omega)}{2\varepsilon+\omega-\Sigma^{R}(\varepsilon+\omega)-\Sigma^{R/A}(\varepsilon)-\omega}
\right].
\end{align} 
After some algebra, the vertex corrections can be expressed as
\begin{align}\label{vertRR}
\frac{v}{v-v_x^{RR}(\varepsilon,\omega)}&=
\frac{3[2\varepsilon+\omega-\Sigma^{R}(\varepsilon+\omega)-\Sigma^{R}(\varepsilon)]
[\omega-\Sigma^{R}(\varepsilon+\omega)+\Sigma^{R}(\varepsilon)]}
{[\varepsilon-\Sigma^{R}(\varepsilon)][3\omega-4\Sigma^{R}(\varepsilon)]
+[\varepsilon+\omega-\Sigma^{R}(\varepsilon+\omega)][3\omega-4\Sigma^{R}(\varepsilon+\omega)]},
\\
\label{vertRA}
\frac{v}{v-v_x^{RA}(\varepsilon,\omega)}
&=\frac{3[2\varepsilon+\omega-\Sigma^{R}(\varepsilon+\omega)-\Sigma^{A}(\varepsilon)]
[\omega-\Sigma^{R}(\varepsilon+\omega)+\Sigma^{A}(\varepsilon)]}
{[\varepsilon-\Sigma^{A}(\varepsilon)][3\omega-4\Sigma^{A}(\varepsilon)]
+[\varepsilon+\omega-\Sigma^{R}(\varepsilon+\omega)][3\omega-4\Sigma^{R}(\varepsilon+\omega)]}.
\end{align}
In the main text, these results are used in the explicit formula (\ref{cond-explicit}) expressing
the conductivity through the self-energies for an arbitrary disorder strength.

\end{widetext}


%

\end{document}